# Exact Solution of the *N*-dimensional Radial Schrödinger Equation via Laplace Transformation Method with the Generalized Cornell Potential

M. Abu-Shady[1], T. A. Abdel-Karim[1], E. M. Khokha[2]

Department of Applied Mathematics, Faculty of Science, Menoufia University, Shebin El- Kom, Egypt[1]
Department of Basic Science, Modern Academy for Engineering and Technology, Cairo, Egypt[2]

**Abstract**

The exact solution of *N*- dimensional radial Schrödinger equation with the generalized Cornell potential has been obtained using the Laplace transformation (LT) method. The energy eigenvalues and the corresponding wave functions for any state have been determined. The eigenvalues for some special cases of the generalized Cornell potential are obtained. The present results are applied to calculate the mass spectra of heavy quarkonium systems such as charmonium and bottomonium and the $b\bar{c}$ meson. A comparison is discussed with the experimental data and recent other works. The present results are improved in comparison with other recent studies and are in a good agreement with the experimental data. The effect of the dimensional number (*N*) on the meson masses has been studied. We note that the meson masses increase in higher dimensions.

**Keywods:** Schrödinger equation, Laplace transformation, Cornell potential, heavy quarkonium systems

## Introduction

The exact solution of the Schrödinger equation (SE) with spherically symmetric potential plays a vital role in nuclei, atoms, molecules, spectroscopy and in many fields of modern physics. Therefore, the authors have interested to obtain the exact solution of Schrödinger equation using different methods such as point canonical transformation (PCT) [1], Hill determinant method (HDM) [2] and numerical methods [3-5].

Recently, many researchers have shown great attempts for solving the Schrödinger equation in higher dimensions. The study of the higher dimensions is more general and it can directly obtain the results in lower dimensions.

The *N*-dimensional Schrödinger equation has been solved with various types of spherically symmetric potential as the Hydrogen atom [6], Coulomb potential [7], the harmonic oscillator [8], Pseudoharmonic potential [9], Mie-type potential [10], forth-order inverse power potential [11], energy dependent potential [12], Cornell potential [13], quark-antiquark interaction potential [14] and the extended Cornell potential [15].

There are numerous alternative methods in the literature to find the exact solution of the *N*-dimensional Schrödinger equation as supersymmetric quantum mechanics (SUSQM) **[2]**, the Nikiforov-Uvarov (NU) method **[12, 16 - 18]**, power series technique (PST) **[19]**, the asymptotic iteration method (AIM) **[20]**, Pekeris type approximation (PTA) **[19, 21]** and the analytical exact iteration method (AEIM) **[15]**.

One of the most available methods in the literature that has successed to give the exact solution of Schrödinger equation is Laplace transformation method. The Laplace transformation method depends on reducing the second order differential equation into a first order differential equation. The Laplace transformation method has been used in a first time by Schrödinger on the quantum mechanical hydrogen atom in his first paper **[22]** then Swainson and Darke have used the Laplace transformation method to calculate the radial wave functions for the hydrogen atom **[23]**. The one dimensional independent Schrödinger equation has been solved with Laplace transformation method with assuming the potential function is real and expressing it as a Fourier series **[24]** and with Morse potential **[25]**. The *N*-dimensional Schrödinger equation has been investigated via Laplace transformation method with the Coulomb potential **[7]** and with the harmonic oscillator **[26]**. Arda and Sever **[27]** employed Laplace transformation method to derive the exact solution of one dimensional Schrödinger equation with the Morse potential **[27]** and calculate the exact solution of 3-dimensional Schrödinger equation with the Pseudoharmonic potential and Mie-type potentials **[28]** and also to find the exact solutions for some non-central potential **[29]**. The one dimensional quantum harmonic oscillator has been examined via Laplace transformation method in **[30]**. The analytical solution of *N*-dimensional Schrödinger equation has been investigated with Morse potential via Laplace transformation in **[31]**. In Ref. **[32]**, the authors have shown a universal Laplace transformation approach to solve the Schrödinger equation for all known solvable models. The Pseudoharmonic potential **[33]** and the Mie-type potential **[34]** have been investigated via Laplace transformation within the *N*-dimensional Schrödinger equation. Das **[35]** has employed the Laplace transformation at the singular point of the differential equation to derive the exact solution of *N*-dimensional Schrödinger equation for anharmonic potential. On the other hand, many efforts have been introduced to calculate the mass spectra of heavy quarkonium systems such as charmonium and bottomonium using (SE) as in Refs. **[14-15, 17, 20, 36-38]**.

In this paper, we have introduced the generalized Cornell potential which takes the following form:

$$V(r) = a_1 r^2 + a_2 r - \frac{a_3}{r} - \frac{a_4}{r^2} + a_5 \tag{1}$$

So far, no attempts to study this potential using Laplace transformation method. Special cases are obtained from the present potential, Moreover; the present potential is applied on quarkonium properties which gives a good agreement with experimental data.

The paper is organized as follows: The background of the study of previous efforts is introduced in the section 1. In Sec. 2, overview of Laplace transform method is shown. In Sec.3, the exact solution of the $N$-dimensional radial Schrödinger equation is derived. In Sec.4, the results are discussed. In Sec. 5, summary and conclusion are presented.

## 2. Overview of Laplace Transform Method

The Laplace transform $\phi(z)$ or $\mathcal{L}$ of a function $f(t)$ is defined by [39].

$$\phi(z) = \mathcal{L}\{f(t)\} = \int_0^\infty e^{-zt} f(t) dt. \qquad (2)$$

If there is some constant $\sigma \in R$ such that $|e^{-\sigma t} f(t)| \leq M$ for sufficiently large $t$, the integral in equation (2) exist for $\operatorname{Re} z > \sigma$. The Laplace transform may fail to exist because of a sufficiently strong singularity in the function $f(t)$ as $t \to 0$. In particular

$$\mathcal{L}\left[\frac{t^\alpha}{\Gamma(\alpha+1)}\right] = \frac{1}{z^{\alpha+1}}, \alpha > -1 \qquad (3)$$

The Laplace transform has the derivative properties

$$\mathcal{L}\{f^{(n)}(t)\} = z^n \mathcal{L}\{f(t)\} - \sum_{k=0}^{n-1} z^{n-1-k} f^{(k)}(0), \qquad (4)$$

$$\mathcal{L}\{t^n f(t)\} = (-1)^n \phi^{(n)}(z), \qquad (5)$$

where the superscript $(n)$ stands for the $n$-th derivative with respect to $t$ for $f^{(n)}(t)$, and with respect $z$ to for $\phi^{(n)}(z)$. If $z_0$ is the singular point, the Laplace transform behaves $z \to z_0$ as

$$\phi(z) = \frac{1}{(z-z_0)^\upsilon}, \qquad (6)$$

Then for $t \to \infty$

$$f(t) = \frac{1}{\Gamma(\upsilon)} t^{\upsilon-1} e^{z_0 t}, \qquad (7)$$

Where $\Gamma(\upsilon)$ is he gamma function. On the other hand, if near origin $f(t)$ behaves like $t^\alpha$ with $\alpha > -1$, then $\phi(z)$ behaves near $z \to \infty$ as

$$\phi(z) = \frac{\Gamma(\alpha+1)}{z^{\alpha+1}}. \tag{8}$$

## 3. Exact Solution of the *N*-dimensional Radial Schrödinger Equation with the Generalized Cornell Potential

The *N*-dimensional radial Schrödinger equation for two particles interacting via symmetric potential (1) takes the form [19].

$$\left[\frac{d^2}{dr^2} + \frac{(N-1)}{r}\frac{d}{dr} - \frac{\ell(\ell+N-2)}{r^2} + 2\mu(E - V(r))\right]\Psi(r) = 0. \tag{9}$$

Where $\ell, N$ denote the angular quantum number, the dimensional number respectively, and $\mu = \frac{m_1 m_2}{m_1 + m_2}$ is the reduced mass of the two particles.

Substituting from Eq. (1) then

$$\left[\frac{d^2}{dr^2} + \frac{(N-1)}{r}\frac{d}{dr} - \frac{\ell(\ell+N-2)}{r^2} + \varepsilon - A_1 r^2 - A_2 r + \frac{A_3}{r} + \frac{A_4}{r^2} - A_5\right]\Psi(r) = 0. \tag{10}$$

where $\varepsilon = 2\mu E, A_1 = 2\mu a_1, A_2 = 2\mu a_2, A_3 = 2\mu a_3, A_4 = 2\mu a_4,$ and $A_5 = 2\mu a_5.$ (11)

The complete solution of Eq. (10) takes the form

$$\Psi(r) = r^k e^{-\alpha r^2} f(r), k > 0, \text{ with } \alpha = \sqrt{\frac{\mu a_1}{2}}. \tag{12}$$

where the term $r^k$ assures that, the solution at $r = 0$ is bounded. The function $f(r)$ yet to be determined. From Eq. (12) we get

$$\Psi'(r) = r^k e^{-\alpha r^2}\left[f'(r) + \left(\frac{k}{r} - 2\alpha r\right)f(r)\right]. \tag{13}$$

$$\Psi''(r) = r^k e^{-\alpha r^2}\left\{f''(r) + \left(\frac{2k}{r} - 4\alpha r\right)f'(r) + \left[\frac{k(k-1)}{r^2} + 4\alpha^2 r^2 - 4\alpha k - 2\alpha\right]f(r)\right\}. \tag{14}$$

Substituting from Eqs. (12), (13) and (14) into Eq. (10) yields

$$r f''(r) + (\omega - 4\alpha r^2)f'(r) + \left\{\frac{\lambda}{r} - A_2 r^2 + \zeta r + A_3\right\}f(r) = 0. \tag{15}$$

Where,

$$\omega = 2k + N - 1. \tag{16}$$

$$\lambda = k(k + N - 2) - \ell(\ell + N - 2) + A_4. \tag{17}$$

$$\zeta = \varepsilon - 4\alpha k - 2\alpha N - A_5. \tag{18}$$

In order to get the Laplace transform of the above differential equation, we put the parametric condition [33, 35].

$$\lambda = 0. \tag{19}$$

Thus, Eq. (17) has a solution

$$k = \frac{(2-N) \pm \sqrt{(N+2\ell-2)^2 - 4A_4}}{2}. \tag{20}$$

The positive sign in Eq. (20) is taken as in Refs. [33, 35].

Substituting from Eq. (19) into Eq. (15) yields

$$r f''(r) + (\omega - 4\alpha r^2) f'(r) + [\zeta r - A_2 r^2 + A_3] f(r) = 0. \tag{21}$$

Laplace transform of Eq. (21). Identifying the Laplace transform $\phi(z) = \mathcal{L}\{f(r)\}$ and taking the boundary condition $f(0) = 0$, yields:

$$(z+\tau)\frac{d^2\phi(z)}{dz^2} + \left(\frac{z^2}{4\alpha} + \delta\right)\frac{d\phi(z)}{dz} + \left(\rho z - \frac{A_3}{2\alpha}\right)\phi(z) = 0. \tag{22}$$

where,

$$\tau = \frac{\mu a_2}{2\alpha}; \quad \delta = \frac{1}{4\alpha}(\varepsilon - A_5) - \left(k - 2 + \frac{N}{2}\right); \quad \rho = \frac{1}{4\alpha}(3 - N - 2k). \tag{23}$$

The singular point of Eq. (22) is $z = -\tau$. By using the condition of Eq. (6) the solution of Eq. (22) takes the form

$$\phi(z) = \frac{C}{(z+\tau)^{n+1}}, n = 0, 1, 2, 3, \ldots. \tag{24}$$

From Eq. (24), we get

$$\phi'(z) = \frac{-C(n+1)}{(z+\tau)^{n+2}}, \tag{25}$$

$$\phi''(z) = \frac{C(n+1)(n+2)}{(z+\tau)^{n+3}}. \tag{26}$$

Substituting from Eqs. (24), (25) and (26) into Eq. (22) we obtain the following relations

$$\rho = \frac{n+1}{4\alpha}, \tag{27}$$

$$\tau = \frac{A_3}{4\alpha\rho}, \tag{28}$$

$$(n+1)(n-\delta+2) = \frac{A_3\tau}{4\alpha}. \tag{29}$$

Using Eqs. (11), (23) and the set of Eqs. (27-29), then, the energy eigenvalue of Eq. (9) in the $N$-dimensional is given by the relation

$$E_{n\ell N} = \sqrt{\frac{a_1}{2\mu}} \left[ 2n+2+\sqrt{(N+2\ell-2)^2 - 8\mu a_4} \right] - \frac{a_2^2}{4a_1} + a_5. \tag{30}$$

Now, taking the inverse Laplace transform such that $f(r) = \mathcal{L}^{-1}\{\phi(z)\}$. The function $f(r)$ takes the following form

$$f(r) = \frac{C}{\Gamma(n+1)} r^n e^{-\tau r}. \tag{31}$$

Using Eqs. (11), (13) and (23) the eigenfunctions of Eq. (9) are take the following form

$$\Psi(r) = \frac{C}{\Gamma(n+1)} r^{n+k} \exp\left( -\sqrt{\frac{\mu a_1}{2}} r^2 - \sqrt{\frac{\mu}{2a_1}} a_2 r \right). \tag{32}$$

The normalization constant $C$ can be determined by the condition $\int_0^\infty |\Psi(r)|^2 r^{N-1} dr = 1$.

## 4. Discussion of Results

In this section, the energy eigenvalues have been derived for several potentials which used in earlier works and then, the mass spectra of heavy quarkonium systems such as charmonium and bottomonium and the $b\bar{c}$ meson have been calculated.

## 4.1 Special Cases of the Generalized Cornell Potential

### I. Extended Cornell potential

Putting $a_5 = 0$ in Eq. (1) then the extended Cornell potential is obtained as

$$V(r) = a_1 r^2 + a_2 r - \frac{a_3}{r} - \frac{a_4}{r^2} \tag{33}$$

The energy eigenvalues are:

$$E_{n\ell N} = \sqrt{\frac{a_1}{2\mu}} \left[ 2n + 2 + \sqrt{(N + 2\ell - 2)^2 - 8\mu a_4} \right] - \frac{a_2^2}{4a_1}. \tag{34}$$

This result agrees with the result in **[15]**. For *N*=3 in Eq. (34) we have

$$E_{n\ell}^{N=3} = \sqrt{\frac{a_1}{2\mu}} \left[ 2n + 2 + \sqrt{(2\ell + 1)^2 - 8\mu a_4} \right] - \frac{a_2^2}{4a_1}. \tag{35}$$

This result as obtained in **[40]**.

### II. Cornell plus harmonic potential

In this case $a_4 = 0$ and $a_5 = 0$ in Eq. (1) then the Cornell plus harmonic potential is given by the relation

$$V(r) = a_1 r^2 + a_2 r - \frac{a_3}{r}. \tag{36}$$

The energy eigenvalues of the Cornell plus harmonic potential are

$$E_{n\ell N} = \sqrt{\frac{a_1}{2\mu}} (2n + 2\ell + N) - \frac{a_2^2}{4a_1}. \tag{37}$$

This result is closed with result in Ref **[35]**. For *N*=3 in Eq. (37) we obtain

$$E_{n\ell}^{N=3} = \sqrt{\frac{a_1}{2\mu}} (2n + 2\ell + 3) - \frac{a_2^2}{4a_1}. \tag{38}$$

The result in Eq. (38) has been obtained in **[41]**.

### III. Pseudoharmonic potential

For $a_2 = 0$, $a_3 = 0$ the Pseudoharmonic potential having the form

$$V(r) = a_1 r^2 - \frac{a_4}{r^2} + a_5. \tag{39}$$

The energy spectrum of the Pseudoharmonic potential is obtained:

$$E_{nl}^N = \sqrt{\frac{8a_1}{\mu}} \left[ \frac{n}{2} + \frac{1}{2} + \frac{1}{4}\sqrt{(N+2\ell-2)^2 - 8\mu a_4} \right] - \frac{a_2^2}{4a_1} + a_5. \tag{40}$$

The obtained results are in good agreement with Ref [33]. For $N=3$ in Eq. (39), we get

$$E_{n\ell}^{N=3} = \sqrt{\frac{8a_1}{\mu}} \left[ \frac{n}{2} + \frac{1}{2} + \frac{1}{4}\sqrt{(2\ell+1)^2 - 8\mu a_4} \right] - \frac{a_2^2}{4a_1} + a_5. \tag{41}$$

Eq. (41) show a qualitative agreement with the result is in [28].

### IV. Isotopic harmonic plus inverse quadratic potential

Assuming $a_1 = \frac{\mu\omega^2}{2}, a_4 = -g$ and $a_2 = a_3 = a_5 = 0$ the isotopic harmonic plus inverse quadratic potential take the form

$$V(r) = \frac{\mu\omega^2}{2} r^2 + \frac{g}{r^2}. \tag{42}$$

The eigenvalues of the bound states at $N=3$ are:

$$E_{n\ell}^{N=3} = \frac{\omega}{2} \left[ 2n + 2 + \sqrt{(2\ell+1)^2 + 8\mu g} \right]. \tag{43}$$

This result is in a good agreement with Ref [42].

### V. Isotopic harmonic oscillator quadratic potential

We take $a_1 = \frac{\mu\omega^2}{2}$ and $a_2 = a_3 = a_4 = a_5 = 0$ then, the isotopic harmonic plus inverse quadratic potential can be written as

$$V(r) = \frac{\mu\omega^2}{2} r^2. \tag{44}$$

The energy eigenvalues becomes:

$$E_{n\ell N} = \omega(n + \ell + N/2). \tag{45}$$

This result is very much similar to the ones obtained in **[26, 42]**. For *N*=3 in Eq. (45) leads

$$E_{n\ell}^{N=3} = \omega\left(n+\ell+\tfrac{3}{2}\right). \quad (46)$$

This results show agreement in comparison with Ref. **[42]**.

### 4.2. Mass Spectra of Heavy Quarkonia and $b\bar{c}$ meson masses

Here, the mass spectra of heavy quarkonia such as charmonuim, bottomonium and $b\bar{c}$ meson have been calculated using the generalized Cornell potential (1). The following formula **[38]** has been used for determining quarkonium masses.

$$M = m_1 + m_2 + E_{nl}^N. \quad (47)$$

Substituting from Eq. (30) into Eq. (47), the in the mass spectra of charmonium can be given by

$$M = 2m_c + \sqrt{\frac{a_1}{2\mu}}\left[2n+2+\sqrt{(N+2\ell-2)^2-8\mu a_4}\right]-\frac{a_2^2}{4a_1}+a_5. \quad (48)$$

Also, the mass spectra of bottomonium can be calculated from Eq. (48) by replacing $m_c$ by $m_b$

**Table 1:** Mass spectra of charmonium in (GeV)
$m_c = 1.209$ GeV, $a_1 = 0.067298$ GeV³, $a_2 = 0.09$ GeV², $a_4 = 0.0016$ GeV⁻¹, $a_5 = 0.00148$ GeV.

| State | Present work | [14] | [15] | Exp. | N=4 | N=5 |
|---|---|---|---|---|---|---|
| 1S | 3.0963 | 3.078 | 3.0955 | 3096 | 3.3327 | 3.5687 |
| 1P | 3.5687 | 3.415 | 3.568 | --- | 3.8048 | 4.0407 |
| 1D | 4.0407 | 3.581 | 4.0397 | --- | 4.2767 | 4.5127 |
| 2S | 3.5681 | 3.749 | 3.5674 | 3.686 | 3.8045 | 4.0406 |
| 2P | 4.0406 | 3.917 | 4.0396 | 3.773 | 4.2766 | 4.5126 |
| 3S | 4.0400 | 4.085 | 4.0392 | 4.040 | 4.2764 | 4.5125 |
| 4S | 4.5119 | 4.589 | 4.5111 | 4.263 | 4.7483 | 4.9843 |

**Table 2**: Mass spectra of bottomonium in (GeV)

$m_b = 4.823$ GeV, $a_1 = 0.09344$ GeV$^3$, $a_2 = 0.34458$ GeV$^2$, $a_4 = 0.0015$ GeV$^{-1}$, $a_5 = 0.001$ GeV.

| State | Present work | [14] | [15] | Exp. | N=4 | N=5 |
|---|---|---|---|---|---|---|
| 1S | 9.7449 | 9.510 | 9.7447 | 9.460 | 9.8851 | 10.0246 |
| 1P | 10.0246 | 9.862 | 10.0241 | --- | 10.1639 | 10.3032 |
| 1D | 10.3032 | 10.038 | 10.3027 | --- | 10.4425 | 10.5817 |
| 2S | 10.0232 | 10.214 | 10.0232 | 10.023 | 10.1635 | 10.3029 |
| 2P | 10.3029 | 10.390 | 10.3025 | --- | 10.4423 | 10.5816 |
| 3S | 10.3016 | 10.566 | 10.3016 | 10.355 | 10.4418 | 10.5814 |
| 4S | 10.5800 | 11.094 | 10.5800 | 10.580 | 10.7202 | 10.8597 |

**Table 3**: Mass spectra of $b\bar{c}$ meson in (GeV)

$m_b = 4.823$ GeV, $m_c = 1.209$ GeV, $a_1 = 0.279294$ GeV$^3$, $a_2 = 1.00482$ GeV$^2$, $a_4 = 0.001$ GeV$^{-1}$, $a_5 = 0.01$ GeV.

| State | present work | [43] | [15] | Exp. | N=4 | N=5 |
|---|---|---|---|---|---|---|
| 1S | 6.2770 | 6.349 | 6.2775 | 6.277 | 6.6578 | 7.0381 |
| 1P | 7.0381 | 6.715 | 7.0386 | --- | 7.4183 | 7.7985 |
| 2S | 7.0372 | 6.821 | 7.0374 | --- | 7.4179 | 7.7983 |
| 2P | 7.7983 | 7.102 | 7.7988 | --- | 8.1785 | 8.5586 |
| 3S | 7.7973 | 7.175 | 7.7978 | --- | 8.1781 | 8.5585 |

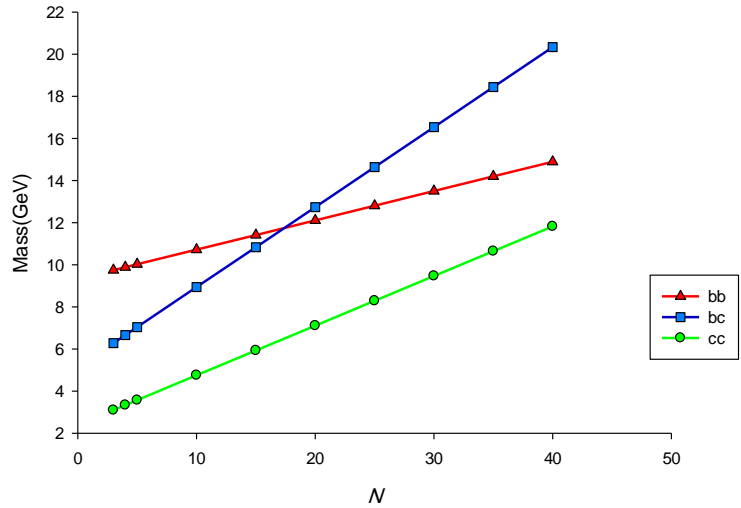

**Figure 1.** The mass spectra of charmonium, bottomonium and $b\bar{c}$ meson with the dimensional number (*N*).

The mass spectra of charmonium, bottomonium and $b\bar{c}$ meson are given in **Table 1**, **Table 2**, and **Table 3**, respectively in comparison with the experimental data and other theoretical studies **[14-15, 43]** at *N*=3 and also at higher dimensional space at *N*=4 and *N*=5. The values of free parameters $a_1, a_2, a_4$ and $a_5$ are fitted with Eq. (48) and with the experimental data. The present results are in good agreement with the experimental data and are improved in comparison with the other theoretical studies **[14-15, 43]**. The present potential is more general in comparison with the potentials in **[15, 26, 28, 33, 35, 40- 42]**. In Ref. **[14]** the authors have calculated the mass spectra of charmonium and bottomonium using the power series method for the quark–antiquark potential which represents a special case of the present potential at $a_4 = a_5 = 0$. In Ref. **[15]** the authors have employed the analytical exact iteration method (AEIM) for computing the mass spectra of charmonium, bottomonium and $b\bar{c}$ meson for the extended Cornell potential which considers another special case of the present potential at $a_4 = 0$. In **Table 1**, the mass spectra of charmonium states have been calculated at *N*=3 in comparison with the experimental data and other theoretical studies **[14-15]**. We note that **1S** and **3S** states are closed with experimental data and other states of charmonium are improved in comparison with the results in **[14-15]** and show a good agreement with the experimental data. Also the mass spectra of charmonium states have been calculated at higher dimensions at *N*=4 and *N*=5. We noted that the masses of all states increase with the increasing the dimensional number (*N*). The mass spectra of bottomonium states have been given at *N*=3 in **Table 2** in comparison with the experimental data and with other theoretical studies **[14-15]**. We note that the values of **2S** and **4S** states close with the experimental data. The value of **3S** state differs from the experimental data with relative error about 0.5%. The other states of bottomonium are improved in comparison with the results in **[14-15]** and are good agreement with the experimental data. Additionally, the mass spectra of bottomonium states have been calculated at higher dimensions at *N*=4 and *N*=5. One notes that increasing the dimensional number (*N*) leads to increase the mass spectra of all states of the bottomonium. We have calculated the different masses of $b\bar{c}$ meson at *N*=3 in comparison with the experimental data and other theoretical studies **[15, 43]** in **Table 3**. The **1S** state closes with the experimental data and the other states of the $b\bar{c}$ meson are in a good agreement with the results in **[15, 43]**. In addition, the mass spectra of the $b\bar{c}$ meson have been calculated at higher dimensions at *N*=4 and *N*=5. We can see that the mass spectra of all states of the $b\bar{c}$ meson increase with increasing the dimensional number (*N*). In **Figure 1,** the mass spectra of charmonium, bottomonium and the $b\bar{c}$ meson have been plotted at higher dimensions. We noted that the meson masses increase with the dimensional number (*N*). This behavior is agreed with the recent results of Ref. **[44]**.

## 5. Summary and conclusion

In this study, we have employed the Laplace transformation (LT) method to find the exact solution of the *N*-dimensional Schrödinger equation with the generalized Cornell potential. The energy eigenvalues and the corresponding wave functions have been obtained. We have derived the energy eigenvalues of some special cases from the generalized Cornell potential in the *N*-dimensional space. In addition, we used the present potential to calculate the mass spectra of heavy quarkonium systems such as charmonium, bottomonium and the $b\bar{c}$ meson in comparison with the experimental data and other theoretical studies at *N*=3. Additionally, the effect of the dimensional number (*N*) on the meson masses has been studied. We noted that the meson masses increase with the dimensional number (*N*). Therefore, we conclude that the present potential using (LT) method gives satisfied results in comparison with experimental data and other recent works.